\documentclass[conference]{IEEEtran}
\IEEEoverridecommandlockouts
\usepackage{cite}
\usepackage{amsmath,amssymb,amsfonts}
\usepackage{algorithmic}
\usepackage{graphicx}
\usepackage{textcomp}
\usepackage{xcolor}
\usepackage{tikz}
\usepackage{pgfplots}
\usepackage{tikzscale}
\usepackage[font=small]{subcaption}
\usepackage[font=small]{caption}
\usepackage[acronyms,nonumberlist,nopostdot,nomain,nogroupskip]{glossaries}
\usepackage{booktabs}
\usepackage[draft]{hyperref}
\usepackage[capitalize]{cleveref}

\newacronym{quic}{QUIC}{Quick UDP Internet Connections}
\newacronym{3gpp}{3GPP}{3rd Generation Partnership Project}
\newacronym{adc}{ADC}{Analog to Digital Converter}
\newacronym{5g}{5G}{5th generation}
\newacronym{aimd}{AIMD}{Additive Increase Multiplicative Decrease}
\newacronym{am}{AM}{Acknowledged Mode}
\newacronym{amc}{AMC}{Adaptive Modulation and Coding}
\newacronym{aqm}{AQM}{Active Queue Management}
\newacronym{awgn}{AGWN}{Additive White Gaussian Noise}
\newacronym{balia}{BALIA}{Balanced Link Adaptation}
\newacronym{bdp}{BDP}{Bandwidth-Delay Product}
\newacronym{bf}{BF}{BeamForming}
\newacronym{hbf}{HBF}{Hybrid Beamforming}
\newacronym{cc}{CC}{Congestion Control}
\newacronym{cdf}{CDF}{Cumulative Distribution Function}
\newacronym{pdf}{PDF}{Probability Density Function}
\newacronym{cn}{CN}{Core Network}
\newacronym{cqi}{CQI}{Channel Quality Information}
\newacronym{cp}{CP}{Control Plane}
\newacronym{csirs}{CSI-RS}{Channel State Information - Reference Signal}
\newacronym{dc}{DC}{Dual Connectivity}
\newacronym{dce}{DCE}{Direct Code Execution}
\newacronym{dci}{DCI}{Downlink Control Information}
\newacronym{dl}{DL}{Downlink}
\newacronym{dmr}{DMR}{Deadline Miss Ratio}
\newacronym{dmrs}{DMRS}{DeModulation Reference Signal}
\newacronym{e2e}{E2E}{End-to-End}
\newacronym{ecn}{ECN}{Explicit Congestion Notification}
\newacronym{edf}{EDF}{Earliest Deadline First}
\newacronym{enb}{eNB}{evolved Node Base}
\newacronym{epc}{EPC}{Evolved Packet Core}
\newacronym{es}{ES}{Edge Server}
\newacronym{fdma}{FDMA}{Frequency Division Multiple Access}
\newacronym{fdd}{FDD}{Frequency Division Duplexing}
\newacronym[firstplural=Radio Access Technologies (RATs)]{rat}{RAT}{Radio Access Technology}
\newacronym{fs}{FS}{Fast Switching}
\newacronym{ftp}{FTP}{File Transfer Protocol}
\newacronym{gnb}{gNB}{Next Generation Node Base}
\newacronym{harq}{HARQ}{Hybrid Automatic Repeat reQuest}
\newacronym{hetnet}{HetNet}{Heterogeneous Network}
\newacronym{hh}{HH}{Hard Handover}
\newacronym{hol}{HOL}{Head-of-Line}
\newacronym{ia}{IA}{Initial Access}
\newacronym{imt}{IMT}{International Mobile Telecommunication}
\newacronym{iot}{IoT}{Internet of Things}
\newacronym{los}{LOS}{Line of Sight}
\newacronym{lte}{LTE}{Long Term Evolution}
\newacronym{m2m}{M2M}{Machine to Machine}
\newacronym{mac}{MAC}{Medium Access Control}
\newacronym{mc}{MC}{Multi-Connectivity}
\newacronym{mcs}{MCS}{Modulation and Coding Scheme}
\newacronym{mec}{MEC}{Mobile Edge Cloud}
\newacronym{mi}{MI}{Mutual Information}
\newacronym{mimo}{MIMO}{Multiple-Input Multiple-Output}
\newacronym{mumimo}{MU-MIMO}{Multi-User Multiple-Input Multiple-Output}
\newacronym{mmwave}{mmWave}{millimeter wave}
\newacronym{mr}{MR}{Maximum Rate}
\newacronym{mss}{MSS}{Maximum Segment Size}
\newacronym{mtd}{MTD}{Machine-Type Device}
\newacronym{mtu}{MTU}{Maximum Transmission Unit}
\newacronym{nfv}{NFV}{Network Function Virtualization}
\newacronym{nlos}{NLOS}{Non Line of Sight}
\newacronym{nr}{NR}{New Radio}
\newacronym{ns3}{ns-3}{Network Simulator 3}
\newacronym{ofdm}{OFDM}{Orthogonal Frequency Division Multiplexing}
\newacronym{ofdma}{OFDMA}{Orthogonal Frequency Division Multiple Access}
\newacronym{pdcch}{PDCCH}{Physical Downlink Control Channel}
\newacronym{pucch}{PUCCH}{Physical Uplink Control Channel}
\newacronym{pdcp}{PDCP}{Packet Data Convergence Protocol}
\newacronym{pdsch}{PDSCH}{Physical Downlink Shared Channel}
\newacronym{pdu}{PDU}{Packet Data Unit}
\newacronym{pf}{PF}{Proportional Fair}
\newacronym{pgw}{PGW}{Packet Gateway}
\newacronym{phy}{PHY}{Physical}
\newacronym{pbch}{PBCH}{Physical Broadcast Channel}
\newacronym[plural=\gls{mme}s,firstplural=Mobility Management Entities (MMEs)]{mme}{MME}{Mobility Management Entity}
\newacronym{prb}{PRB}{Physical Resource Block}
\newacronym{pss}{PSS}{Primary Synchronization Signal}
\newacronym{pusch}{PUSCH}{Physical Uplink Shared Channel}
\newacronym{rach}{RACH}{Random Access Channel}
\newacronym{ran}{RAN}{Radio Access Network}
\newacronym{red}{RED}{Random Early Detection}
\newacronym{rf}{RF}{Radio Frequency}
\newacronym{rlc}{RLC}{Radio Link Control}
\newacronym{rlf}{RLF}{Radio Link Failure}
\newacronym{rrc}{RRC}{Radio Resource Control}
\newacronym{rrm}{RRM}{Radio Resource Management}
\newacronym{rr}{RR}{Round Robin}
\newacronym{rs}{RS}{Remote Server}
\newacronym{rsrp}{RSRP}{Reference Signal Received Power}
\newacronym{rss}{RSS}{Received Signal Strength}
\newacronym{rtt}{RTT}{Round Trip Time}
\newacronym{rw}{RW}{Receive Window}
\newacronym{rx}{RX}{Receiver}
\newacronym{sa}{SA}{standalone}
\newacronym{sack}{SACK}{Selective Acknowledgment}
\newacronym{sap}{SAP}{Service Access Point}
\newacronym{sch}{SCH}{Secondary Cell Handover}
\newacronym{scoot}{SCOOT}{Split Cycle Offset Optimization Technique}
\newacronym{sdma}{SDMA}{Spatial Division Multiple Access}
\newacronym{sinr}{SINR}{Signal to Interference plus Noise Ratio}
\newacronym{sm}{SM}{Saturation Mode}
\newacronym{snr}{SNR}{Signal to Noise Ratio}
\newacronym{son}{SON}{Self-Organizing Network}
\newacronym{ss}{SS}{Synchronization Signal}
\newacronym{srs}{SRS}{Sounding Reference Signal}
\newacronym{sss}{SSS}{Secondary Synchronization Signal}
\newacronym{tb}{TB}{Transport Block}
\newacronym{tcp}{TCP}{Transmission Control Protocol}
\newacronym{tdd}{TDD}{Time Division Duplexing}
\newacronym{tdma}{TDMA}{Time Division Multiple Access}
\newacronym{tfl}{TfL}{Transport for London}
\newacronym{tm}{TM}{Transparent Mode}
\newacronym{trp}{TRP}{Transmitter Receiver Pair}
\newacronym{tti}{TTI}{Transmission Time Interval}
\newacronym{ttt}{TTT}{Time-to-Trigger}
\newacronym{tx}{TX}{Transmitter}
\newacronym{ue}{UE}{User Equipment}
\newacronym{ul}{UL}{Uplink}
\newacronym{uml}{UML}{Unified Modeling Language}
\newacronym{um}{UM}{Unacknowledged Mode}
\newacronym{utc}{UTC}{Urban Traffic Control}
\newacronym{vm}{VM}{Virtual Machine}
\newacronym{rsrq}{RSRQ}{Reference Signal Received Quality}
\newacronym{rssi}{RSSI}{Received Signal Strength Indicator}
\newacronym{crs}{CRS}{Cell Reference Signal}
\newacronym{comp}{CoMP}{Coordinated Multi-Point}
\newacronym{cran}{C-RAN}{Cloud \acrlong{ran}}
\newacronym{ca}{CA}{Carrier Aggregation}
\newacronym{cco}{CC}{Carrier Component}
\newacronym{nsa}{NSA}{Non Stand Alone}
\newacronym{embb}{eMBB}{Enhanced Mobility Broadband}
\newacronym{bsr}{BSR}{Buffer Status Report}
\newacronym{srb}{SRB}{Service Radio Bearer}
\newacronym{scm}{SCM}{Spatial Channel Model}
\newacronym{sctp}{SCTP}{Stream Control Transmission Protocol}
\newacronym{mptcp}{MPTCP}{Multi-path TCP}
\newacronym{ietf}{IETF}{Internet Engineering Task Force}
\newacronym{os}{OS}{Operating System}
\newacronym{tls}{TLS}{Transport Layer Security}
\newacronym{rfc}{RFC}{Request for Comments}
\newacronym{http}{HTTP}{HyperText Transfer Protocol}
\newacronym{nat}{NAT}{Network Address Translation}
\newacronym{api}{API}{Application Programming Interface}
\newacronym{rto}{RTO}{Retransmission Timeout}
\newacronym{psc}{PSC}{Public Safety Communication}
\newacronym{rpgm}{RPGM}{Reference Point Group Mobility}
\newacronym{ic}{IC}{Incident Command}
\newacronym{rsu}{RSU}{Road Side Unit}
\newacronym{uav}{UAV}{Unmanned Aerial Vehicle}
\newacronym{iab}{IAB}{Integrated Access and Backhaul}
\newacronym{psd}{PSD}{Power Spectral Density}
\newacronym{mpc}{MPC}{Multi Path Component}
\newacronym{rt}{RT}{Ray Tracer}
\newacronym{aoa}{AoA}{Angle of Arrival}
\newacronym{aod}{AoD}{Angle of Departure}
\newacronym{inr}{INR}{Interference to Noise Ratio}
\newacronym{qd}{QD}{Quasi Deterministic}
\newacronym{wlan}{WLAN}{Wireless Local Area Network}
\newacronym{cad}{CAD}{Computer-aided Design}
\newacronym{ap}{AP}{Access Point}
\newacronym{sta}{STA}{Station}
\newacronym{nrmse}{NRMSE}{Normalized Root Mean Square Error}
\newacronym{ut}{UT}{User Terminal}
\newacronym{bs}{BS}{Base Station}
\newacronym{mmse}{MMSE}{Minimum Mean Squared Error}
\newacronym{gbf}{GBF}{Geometric BeamForming}
\newacronym{cbf}{CBF}{Codebook BeamForming}
\newacronym{fmbf}{FMBF}{Frequency-Flat MMSE BeamForming}
\newacronym{smbf}{SMBF}{Frequency-Selective MMSE BeamForming}
\newacronym{bler}{BLER}{Block Error Rate}
\newacronym{fft}{FFT}{Fast Fourier Transform}
\newacronym{nack}{NACK}{Negative Acknowledgment}
\newacronym{ula}{ULA}{Uniform Linear Array}
\newacronym{upa}{UPA}{Uniform Planar Array}
\newacronym{tmrs}{TMRS}{TDMA mmWave RR Scheduler}
\newacronym{pmrs}{PMRS}{Padded mmWave RR Scheduler}
\newacronym{amrs}{AMRS}{Asynchronous mmWave almost-RR Scheduler}
\newacronym{rb}{RB}{Resource Block}
\newacronym{udp}{UDP}{User Datagram Protocol}
\newacronym{noma}{NOMA}{Non Orthogonal Multiple Access}
\newacronym{dft}{DFT}{Discrete Fourier Transform}
\newacronym{cav}{CAV}{Connected Autonomous Vehicles}
\newacronym{fr2}{FR2}{Frequency Range 2}
\newacronym{svd}{SVD}{Singular Value Decomposition}

\usepackage{tikz}
\usepackage{pgfplots}
\pgfplotsset{compat=newest} 
\pgfplotsset{plot coordinates/math parser=false} 
\newlength\fheight
\newlength\fwidth
\usetikzlibrary{plotmarks,patterns,decorations.pathreplacing,backgrounds,calc,arrows,arrows.meta,spy,matrix}
\usepgfplotslibrary{patchplots,groupplots}
\usepackage{tikzscale}

\tikzstyle{startstop} = [rectangle, rounded corners, minimum width=2cm, minimum height=0.5cm,text centered, draw=black]
\tikzstyle{io} = [trapezium, trapezium left angle=70, trapezium right angle=110, minimum width=3cm, minimum height=1cm, text centered, draw=black]
\tikzstyle{process} = [rectangle, minimum width=2cm, minimum height=0.5cm, text centered, draw=black, align=center]
\tikzstyle{decision} = [ellipse, minimum width=2cm, minimum height=1cm, text centered, draw=black]
\tikzstyle{arrow} = [thick,<->,>=stealth]
\tikzstyle{line} = [thick,>=stealth]
\tikzstyle{darrow} = [thick,<->,>=stealth,dashed]
\tikzstyle{sarrow} = [thick,->,>=stealth]
\tikzstyle{larrow} = [line width=0.1mm,dashdotted,<->,>=stealth]

\pgfplotsset{every tick label/.append style={font=\scriptsize}, 
             every axis/.append style={
             width=\fwidth, height=\fheight, at={(0\fwidth,0\fheight)}, 
             xlabel style={font=\footnotesize\color{white!15!black}},
             xmajorgrids,
             ylabel style={yshift=-0.15cm, font=\footnotesize\color{white!15!black}},
             ymajorgrids,
             legend style={font=\footnotesize\color{white!15!black}},
             /pgfplots/ybar legend/.style={/pgfplots/legend image code/.code={\draw[##1,/tikz/.cd,yshift=-0.25em](0cm,0cm) rectangle (10pt,1em);},},
             }}

\definecolor{SchoolColor}{RGB}{0.71, 0, 0.106}
\definecolor{chaptergrey}{rgb}{0.61, 0, 0.09} 
\definecolor{midgrey}{rgb}{0.4, 0.4, 0.4}
\definecolor{chaptergreen}{rgb}{0.09, 0.612, 0}
\definecolor{chapterpurple}{rgb}{0.522, 0, 0.612}
\definecolor{chapterlightgreen}{rgb}{0, 0.612, 0.522}

\makeatletter
\def\grd@save@target#1{%
  \def\grd@target{#1}}
\def\grd@save@start#1{%
  \def\grd@start{#1}}
\tikzset{
  grid with coordinates/.style={
    to path={%
      \pgfextra{%
        \edef\grd@@target{(\tikztotarget)}%
        \tikz@scan@one@point\grd@save@target\grd@@target\relax
        \edef\grd@@start{(\tikztostart)}%
        \tikz@scan@one@point\grd@save@start\grd@@start\relax
        \draw[minor help lines] (\tikztostart) grid (\tikztotarget);
        \draw[major help lines] (\tikztostart) grid (\tikztotarget);
        \grd@start
        \pgfmathsetmacro{\grd@xa}{\the\pgf@x/1cm}
        \pgfmathsetmacro{\grd@ya}{\the\pgf@y/1cm}
        \grd@target
        \pgfmathsetmacro{\grd@xb}{\the\pgf@x/1cm}
        \pgfmathsetmacro{\grd@yb}{\the\pgf@y/1cm}
        \pgfmathsetmacro{\grd@xc}{\grd@xa + \pgfkeysvalueof{/tikz/grid with coordinates/major step x}}
        \pgfmathsetmacro{\grd@yc}{\grd@ya + \pgfkeysvalueof{/tikz/grid with coordinates/major step y}}
        \foreach \x in {\grd@xa,\grd@xc,...,\grd@xb}
        \node[anchor=north] at (\x,\grd@ya) {\pgfmathprintnumber{\x}};
        \foreach \y in {\grd@ya,\grd@yc,...,\grd@yb}
        \node[anchor=east] at (\grd@xa,\y) {\pgfmathprintnumber{\y}};
      }
    }
  },
  minor help lines/.style={
    help lines,
    gray,
    line cap =round,
    xstep=\pgfkeysvalueof{/tikz/grid with coordinates/minor step x},
    ystep=\pgfkeysvalueof{/tikz/grid with coordinates/minor step y}
  },
  major help lines/.style={
    help lines,
    line cap =round,
    line width=\pgfkeysvalueof{/tikz/grid with coordinates/major line width},
    xstep=\pgfkeysvalueof{/tikz/grid with coordinates/major step x},
    ystep=\pgfkeysvalueof{/tikz/grid with coordinates/major step y}
  },
  grid with coordinates/.cd,
  minor step x/.initial=.5,
  minor step y/.initial=.2,
  major step x/.initial=1,
  major step y/.initial=1,
  major line width/.initial=1pt,
}
\makeatother

\def\BibTeX{{\rm B\kern-.05em{\sc i\kern-.025em b}\kern-.08em
    T\kern-.1667em\lower.7ex\hbox{E}\kern-.125emX}}

\newcommand{\ml}[1]{}
\newcommand{\tz}[1]{}
\newcommand{\sz}[1]{}

\begin{document}

\glsunset{nr}
\glsunset{5g}

\title{A Full-Stack Open-Source Framework for Antenna and Beamforming Evaluation in mmWave 5G NR
\thanks{This work was partially supported by NIST under Award No.~60NANB20D082.
Mattia Lecci's activity was supported by \textit{Fondazione CaRiPaRo} under the grant ``Dottorati di Ricerca 2018.''}
}
\author{\IEEEauthorblockN{Mattia Lecci, Tommaso Zugno, Silvia Zampato, Michele Zorzi}
\IEEEauthorblockA{\textit{Department of Information Engineering}, \textit{University of Padova}, Italy\\
\texttt{\{name.surname@dei.unipd.it\}}}
\vspace*{-1em}
}

\maketitle

\begin{abstract}
  
  \Gls{mmwave} communication represents one of the main innovations of the next generation of wireless technologies, allowing users to reach unprecedented data rates. 
  To overcome the high path loss at \gls{mmwave} frequencies, these systems make use of directional antennas able to focus the transmit power into narrow beams using \gls{bf} techniques, thus making the communication directional. 
  This new paradigm opens up a set of challenges for the design of efficient wireless systems, in which antenna and \gls{bf} components play an important role also at the higher layer of the protocol stack. 
  For this reason, accurate modeling of these components in a full-stack simulation is of primary importance to understand the overall system behavior. 

  This paper proposes a novel framework for the end-to-end simulation of 5G \gls{mmwave} cellular networks, including a ray-tracing based channel model and accurate models for antenna arrays and \gls{bf} schemes.
  We showcase this framework by evaluating the performance of different antenna and \gls{bf} configurations considering both link-level and end-to-end metrics and present the obtained results.
  
  \begin{picture}(0,0)(0,-360)
  \put(0,0){
  \put(0,0){\footnotesize This work has been submitted to the IEEE for possible publication. Copyright may be transferred without notice.}}
  \end{picture}
\end{abstract}


\vspace*{-1em}
\section{Introduction} 
\label{sec:introduction}
\glsunset{3gpp}

One of the focus points of the next generation of wireless networks is the use of the \gls{mmwave} band, often identified with frequencies between 6 and 100~GHz. 
Thanks to the amount of available bandwidth that characterizes this portion of the spectrum, wireless systems that operate at \gls{mmwave} frequencies can use much wider bandwidths compared to legacy technologies. 
As a result, users can obtain unprecedented data rates, in the order of gigabits-per-second, paving the way for new advanced use cases such as virtual and augmented reality.
\gls{3gpp} \gls{nr}~\cite{38300}, the communication standard for the \gls{5g} of cellular networks, currently supports carrier frequencies up to 52.6~GHz, with possible extensions to 71~GHz in future releases~\cite{qualcomm201971}.

\Gls{rf} signals at \gls{mmwave} frequencies, though, present some additional difficulties with respect to their more common sub-6~GHz counterpart~\cite{rangan2014millimeter}.
For example, free space propagation loss increases significantly, making wireless links harder to maintain over long distances.
This, in turn, tends to dissipate high-order reflections and diffractions, making the \gls{mmwave} channel sparser and more directional, reducing both intra-cell and inter-cell interference, thus allowing spatial diversity among different users.
Furthermore, the higher penetration loss occurring in common materials, as well as a deeper diffraction shadow, make it much harder to maintain a reliable connection in \gls{nlos} conditions.
Finally, the shorter wavelength results in a lower channel coherence time in mobile environments, making the connection less stable due to the rapid channel fluctuations.

Using directive antennas, especially in the form of electronically steered phased antenna arrays, can solve most of the problems caused by the physics of the \gls{mmwave} band.
In fact, directivity can compensate for the higher propagation loss by focusing the power into narrow beams, thus restoring the possibility of medium-range transmissions, while also further reducing interference and the contribution of secondary paths to the receiver, thus improving the channel coherence time~\cite{va2015mmwaveCoherence} and the aggregated cell performance.

On the other hand, directional communications make it inherently more difficult to handle mobile scenarios, making it necessary to keep the alignment between the transmitter and receiver beams~\cite{giordani19beamManagement}, introducing more overhead.
Directivity is even more critical when experiencing sudden blockage events.
In this case, the pair of communicating devices should either store backup links or find a new connection on the fly, increasing the overhead and complexity of beam management operations.

The combination of these phenomena may seriously affect the quality experienced by mobile users.
In this context, it is necessary to implement a detailed performance evaluation to identify and address the weakest points of the technology across the full communication stack.
Cellular networks are extremely complex systems, highly correlated with hidden factors that can impact the overall performance at different scales. 
Usually, the different components, i.e., the antenna system, the \gls{rf} components, the cellular protocol stack, etc., are designed using a block-level approach, developing each block independently of the others. 
However, this methodology may lead to undesired behaviors and even sub-optimal performance, since the possible side effects among the components may have a strong impact on the overall system.
This problem can be solved using system-level simulation tools, which represent an accurate and cheap solution to evaluate the overall performance before the actual deployment and to adjust the design of each component accordingly. 
Indeed, simulation enables large scale evaluations, allowing the user to decide the degree of abstraction required by the desired analysis~\cite{zugno19simulation}.

In this work, we showcase an open-source and publicly available~\footnote{https://github.com/signetlabdei/ns3-mmwave-antenna} framework for full-stack 5G NR-compliant simulations, based on the popular open-source \gls{ns3}. 
While other 5G NR simulators already exist, to the best of our knowledge we propose a simulator with (i) a ray-tracing based channel model for mobile users, improving the spatio-temporal coherence over the previous stochastic channel~\cite{zugno2020implementation}, (ii) a flexible antenna module, comprising of multiple parametric antenna elements as well as a generic interface for phased antenna arrays, and (iii) a \gls{bf} module.
This work builds on top of a pre-existing full-stack 5G NR \gls{mmwave} simulator~\cite{mezzavilla2018end}, a popular tool developed jointly by the University of Padova and NYU-Wireless.
Thanks to the integration with \gls{ns3}, this simulator features a detailed implementation of the TCP/IP stack, together with several traffic and mobility models, allowing the community to analyze and compare full-stack behaviors of different physical and protocol setups.

This paper is organized as follows.
In \cref{sec:channel_modeling}, we describe the feature we developed to enable a more realistic modeling of the \gls{mmwave} channel.
In \cref{sec:antenna_arrays}, we present new simulation models for antenna arrays and \gls{bf}. 
In \cref{sec:simulation_setup} we show the simulation setup used to obtained the results for \cref{sec:simulation_results}.
Finally, in \cref{sec:conclusions} we conclude and propose future developments for this work.


\section{Channel Modeling} 
\label{sec:channel_modeling}

When simulating wireless networks, channel modeling plays a key role to make simulations more realistic, and this is even more important when simulating \gls{mmwave} links~\cite{8445856}.
Some key features differentiate a \gls{mmwave} link with respect to a sub-6~GHz link, specifically (i) the higher propagation loss, (ii) the spatial sparsity, (iii) the higher penetration losses and deep diffraction shadows, and (iv) the more prominent diffuse scattering~\cite{lecci2020fullStack}.

While propagation losses are usually based on simple models, the other characteristics make the \gls{mmwave} channel more complicated to both simulate and manage.
For example, spatial sparsity, together with mobile users and directional antennas, require a realistic spatio-temporal correlation to simulate realistic results, especially when beam-tracking techniques need to be evaluated.
Furthermore, higher penetration loss and deep diffraction shadows make \gls{mmwave} suffer from strong blockage due to simple obstacles (e.g., trees, pedestrians, light poles) and even from self-blockage.
Finally, diffuse scattering is one of the main causes of fast and frequency-selective fading, which tends to be significant due to the relatively high surface roughness of building materials with respect to the wavelength of the radiated signal.

While stochastic models have proven to be quite accurate for the lower frequencies\ml{cit}, these effects are more appropriately taken into account by quasi-deterministic approaches, mixing deterministic ray-tracing with stochastic models for diffuse scattering~\cite{lecci20qdFormalization}.

We implemented an add-on for \gls{ns3}, making it possible to read channel traces obtained, for example, by a quasi-deterministic \gls{mmwave} channel simulator.
Both the quasi-deterministic channel simulator~\cite{qd-realization} and the \gls{ns3} add-on~\cite{qd-channel} are available and open source, making it possible for the community to further improve their results with realistic channel modeling, while keeping the simulation complexity under control~\cite{lecci2020fullStack}.


\section{Antenna Arrays and Beamforming} 
\label{sec:antenna_arrays}
\glsreset{bf}
Antennas are a fundamental component of every wireless system.
Given that the size of an antenna is in the order of the wavelength of the desired carrier frequency~\cite{balanis05antenna}, when operating at \gls{mmwave} frequencies antennas sizes are in the order of a few millimeters~\cite{rabbani2015improvement}, thus allowing to design compact antenna arrays with tens or hundreds of elements.

The most common and scalable approach is the use of phased antenna arrays, which allows us to focus the transmission in the desired direction using coherent interference.
By carefully introducing phase shifts to the transmitted signal of each antenna element, it is possible to create complex radiation patterns, a technique called \gls{bf}.
In the simplest case, the overall radiation pattern of the antenna system is computed from the type of antenna element used, the placement of the antenna elements in the 3D space, and the \gls{bf} scheme used.

Typically, antenna and \gls{bf} design is carried out as an independent task, by means of real-world experiments or link-level simulations, without considering it as part of the overall system optimization. 
However, the solutions obtained with this approach may not be able to achieve optimal system-level performance, because they are designed without considering the interactions between the antenna systems and the higher layers of the protocol stack. 

To go beyond this standalone block-level design perspective, new tools able to properly consider all the relevant aspects of the cellular system are required.
For instance, the authors of~\cite{abdelkader2019systemlevel} verified the importance of carrying out system-level simulations for the design of an $8\times8$ hybrid beamformer, since the cross effects between the different system blocks may have a strong impact on the overall performance.
In~\cite{8419177}, the authors present novel antenna array and \gls{bf} solutions for \gls{mmwave} \gls{mimo} systems based on lens antennas, and evaluate the end-to-end performance through system level simulations based on ray-tracing.
Also, in~\cite{menudier2018design} the authors investigate the possibility of co-designing the antennas and the \gls{rf} blocks in the front-end using a system-level platform. 
Although these works tackle antenna and/or \gls{bf} design with a system-level approach, they make use of closed source software or unavailable tools, specifically developed for a single application.

In this work, we propose new models for the end-to-end performance evaluation of antenna and \gls{bf} designs targeted for \gls{mmwave} cellular systems.
Thanks to the integration with \gls{ns3}, these models allow users to evaluate the impact of novel antenna and \gls{bf} solutions on the end-to-end system behavior. 
\cref{sub:antenna_array_model} describes the antenna array model, \cref{sub:antenna_element_model} describes the antenna element model, while \cref{sec:bf-model} describes the \gls{bf} model.

\subsection{Antenna Array Model}
\label{sub:antenna_array_model}

Phased antenna arrays can have extremely diverse geometries, from which their \gls{bf} capabilities are derived.
While it would be possible to create a generic class for arbitrary phased arrays, some geometries (e.g., uniform linear and planar arrays) are extremely popular and deserve specialized methods.
For this reason, we created a generic interface for phased antenna arrays, specifying the polarized element field pattern, the locations of the elements (from which it is possible to compute the phase difference experienced by each antenna element for a transmitting or receiving signal), and the \gls{bf} vector (the phase shifts and amplifications applied to every single element necessary to obtain the desired beam shape).

For this work, we considered the model described in the 3GPP specifications TR~38.901~\cite{3gpp.38.901}.
The standard describes a uniform planar array, meaning that antenna elements are equal and are placed in an equally-spaced $M\times N$ rectangular lattice with vertical spacing $d_V$ and horizontal spacing $d_H$, which form a panel.
In our implementation we consider the simpler case of vertically polarized elements and only a single-panel configuration.

\subsection{Antenna Element Model}
\label{sub:antenna_element_model}

Phased antenna arrays are composed of multiple antenna elements capable of radiating and receiving electromagnetic signals.
Every antenna element has a specific radiation and polarization pattern due to its specific design.
Different antennas are needed in different contexts, e.g., directional elements can be used in multi-sector devices (e.g., \glspl{gnb}), while quasi-isotropic antennas may be used for devices with no preferred communicating direction (e.g., \glspl{ut} with single antenna array).

A huge number of antenna element designs exist in practice, leading us to creating a generic interface allowing users to add their own antenna models. 
In general, it is possible to create antenna elements with pattern measured from real devices to further increase the simulation accuracy.
For this paper, we implemented three of the most common antenna element models, with directivity pattern in dBi $D_{\rm dB}$ in the $\theta$ (inclination) and $\phi$ (azimuth) directions:
\begin{itemize}
  \item \textit{Isotropic antenna element}
  \[
    D_{\rm dB} (\theta, \phi) = 0 
  \]
  \item \textit{3GPP antenna element}~\cite{3gpp.38.901}
  \[
    \begin{aligned}
      D_{\rm v, dB}(\theta) &= -\min \left\{ 12 \left( \frac{\theta-90^\circ}{\theta_{3 \rm dB}} \right)^2, SLA_V \right\}\\
      D_{\rm h, dB}(\phi) &= -\min \left\{ 12 \left( \frac{\phi}{\phi_{3 \rm dB}} \right)^2, A_{\rm max} \right\}\\
      D_{\rm dB} (\theta, \phi) &=  G_{E,\rm max} -\\
          &\min \left\{ -(D_{\rm v, dB}(\theta) +  D_{\rm h, dB}(\phi)), A_{\rm max} \right\}
    \end{aligned}
  \]
  where the side-lobe attenuation in the vertical direction $SLA_V=30$~dB, the maximum attenuation $A_{\rm max}=30$~dB, the vertical and horizontal 3~dB beamwidths are respectively $\theta_{3\rm dB} = \phi_{3\rm dB} = 65^\circ$, and the maximum directional gain of the antenna element is $G_{E,\rm max}=8$~dBi.
  \item \textit{Cosine antenna element}
  \[
    \begin{aligned}
    D_{\rm dB} (\theta, \phi) &= G_{\rm max} +\\
            & 20\log_{10} \left( \cos^{\alpha_h} \left( \frac{\phi}{2} \right) \cos^{\alpha_v} \left( \frac{90^\circ - \theta}{2} \right) \right),
    \end{aligned}
  \]
  where the exponents $\alpha_{h/v}$ can be computed from the beamwidths $\rm BW_{h/v}$ as $\alpha_{h/v} = \frac{-3}{20\log_{10} \cos \frac{\rm BW_{h/v}}{4}}$, and the maximum gain $G_{\rm max}$ can be computed from the directivity formula found in~\cite{balanis05antenna}.
\end{itemize}

\subsection{Beamforming Model}
\label{sec:bf-model}

Multiple \gls{bf} architectures exist, which are commonly divided into three main categories, namely analog, digital, and hybrid.
In analog architectures, a network of phase shifters is used to connect the antenna elements to a single \gls{rf} chain, enabling a passive control of the beam by acting on the elements' phases.
In digital architectures, instead, each antenna element is connected to an independent \gls{rf} chain to provide digital control of the \gls{bf} using baseband processing.
The presence of multiple \gls{rf} chains enables \gls{mumimo} operations, i.e., independent data streams can be transmitted and received simultaneously, possibly serving multiple users at the same time.
Finally, hybrid architectures represent a middle ground between analog and digital approaches, in which the array is divided into multiple sections, each including multiple elements connected to an independent \gls{rf} chain.

Although digital and hybrid architectures have the potential to achieve higher spectral efficiencies, several technological and economic issues still make analog \gls{bf} a valuable choice, also due to its relatively low complexity.
For this reason, in this work we consider the analog architecture and leave the study of the other two categories as future work.

Analog \gls{bf} is achieved by controlling amplitude and phase shift of each antenna element of the phased array; this corresponds to assigning a complex number to each element, which is often identified as a \textit{\gls{bf} vector}.
Several algorithms exist to compute such vectors, each affecting the directivity pattern in a unique way.
Some try to maximize the gain in given directions, some try to suppress side lobes, some try to regulate the beamwidth, some others try to optimize the performance for a given channel estimate, and some others even try to also take into account the interference generated to other users.

In general, two main approaches exist: those based on a channel estimate, and those based on \gls{bf} codebooks.

For the first approach, we implemented an algorithm originally proposed in~\cite{zhang2017ns3} based on the \gls{mimo} \textit{Maximum Ratio Transmission} scheme, in which the optimal weight vectors corresponds to the singular vectors associated with the largest singular values of the \gls{svd} of the estimated channel matrix.
For a perfect channel estimate in interference-free environments, this method ensures optimal performance.
Unfortunately, good channel estimates are hard and expensive to obtain, especially when dealing with large antenna arrays. The \gls{svd} decomposition is itself an expensive operation, and sending feedback information comprises a difficult trade-off between accuracy and overhead.
For this work, we assume that the channel matrix is perfectly known, thus posing the \gls{bf} algorithm in ideal conditions.

For the second approach, we implemented a generic interface for codebooks, allowing the user to create custom ones.
We also implemented a file-based codebook, allowing to create complex codebooks using custom and highly-specialized software, avoiding the computation of sophisticated algorithms on \gls{ns3}.
As a first step, the implemented codebook-based \gls{bf} computes the \gls{sinr} for every pair of TX/RX \gls{bf} vector, choosing the pair with the best performance.
The advantages over the previous approach are many, in particular, no channel estimation nor complex matrix decomposition has to be performed and the only feedback needed is the index corresponding to the best performing \gls{bf} codeword. \ml{cit?}
On the other hand, exhaustive search among all possible codeword pairs may be inefficient, while reducing the search to a subset of codewords might yield sub-optimal performance.
We leave a more realistic and standard-compliant beam-management implementation and evaluation as future work.
\begin{figure}
  \centering
    \setlength\fwidth{0.6\columnwidth} 
    \setlength\fheight{0.6\columnwidth}
    \input{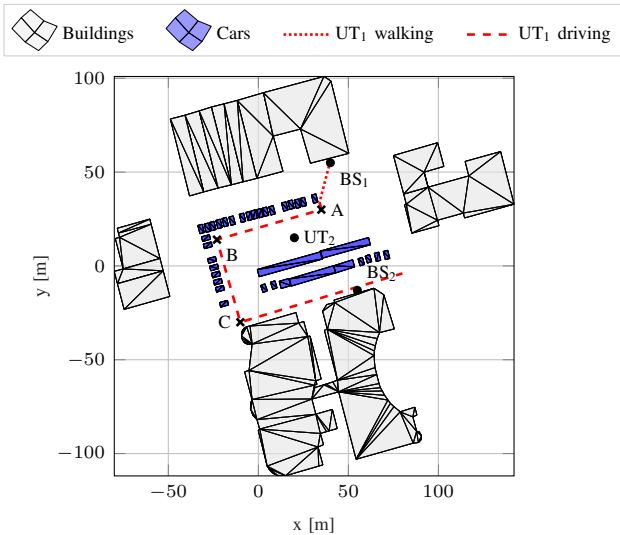}
  \caption{Reference scenario.}
  \label{fig:scenario}
\end{figure}
\begin{figure}
  \centering
  \setlength\fwidth{\columnwidth}
  \setlength\fheight{0.5\columnwidth}
  \input{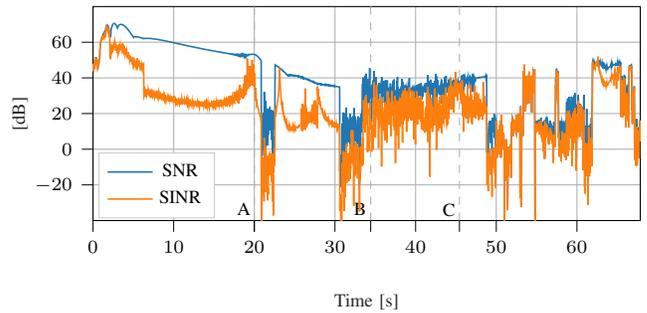}
  \caption{Temporal evolution of the signal strength experienced by UT\textsubscript{1}.
  \ml{why the spike around 24s?}}
  \label{fig:temporal_evolution}
\end{figure}

\begin{table}
  \vspace*{2em}
  \caption{Simulation parameters.}
  \label{table:params}
  \centering
  \begin{tabular}{l|l}
  \toprule
    Frequency & 28 GHz\\
    Bandwidth & 400 MHz \\ 
    Channel sampling period & 5 ms\\
    \gls{nr} numerology index & 2\\
    Transmission power & 30~dBm\\
    Noise figure & 9 dB\\
    \gls{bs} array size & $8\times 8$\\
    \gls{ut} array size & \{$1\times 4$, $4\times 1$, $4\times 4$\}\\
    \gls{bs} element pattern & \{Isotropic, 3GPP, Cosine\}\\
    \gls{ut} element pattern & Isotropic\\
    \gls{bf} algorithm & \{SVD, Codebook\}\\
    Codebook \gls{bf} period & \{10, 100, 1000\} ms\\
    APP packet size & 1490 bytes\\
    Inter-packet interval & \{$10, 1000$\} $\mu$s\\
    RLC mode & Acknowledge Mode (AM)\\
  \bottomrule
  \end{tabular}
  \vspace{-0.5cm}
\end{table}

\begin{figure*}
\setlength{\abovecaptionskip}{-0.1cm}
\setlength\fwidth{\columnwidth}
\setlength\fheight{.5\columnwidth}
\begin{subfigure}[t]{0.48\textwidth}
  \centering
  \input{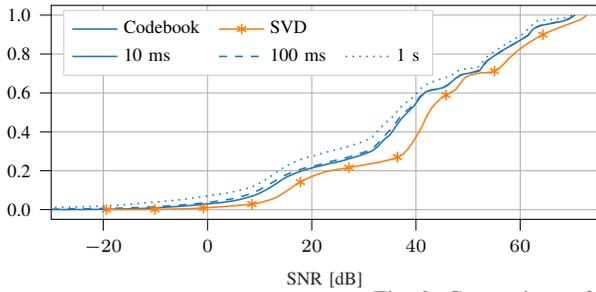}
\end{subfigure}%
\hfill%
\begin{subfigure}[t]{0.48\textwidth}
  \centering
  \input{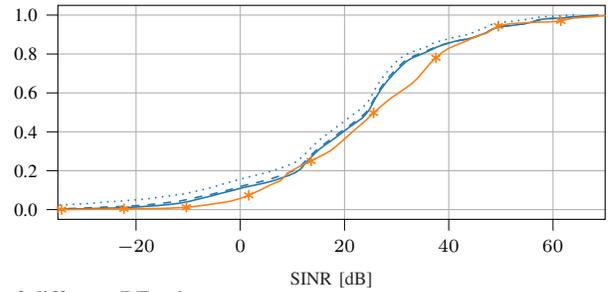}
  \label{fig:bf-comparison-sinr}
\end{subfigure}
\caption{Comparison of the CDFs of different BF schemes.}
\label{fig:bf-comparison}
\end{figure*}

\begin{figure*}
\setlength{\belowcaptionskip}{-0.33cm}
\setlength\fwidth{\columnwidth}
\setlength\fheight{.5\columnwidth}
\begin{subfigure}[t]{0.48\textwidth}
  \centering
  \input{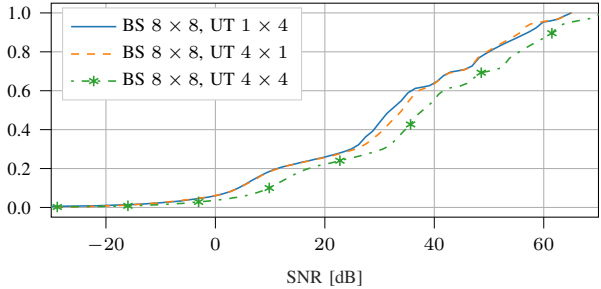}
\end{subfigure}%
\hfill%
\begin{subfigure}[t]{0.48\textwidth}
  \centering
  \input{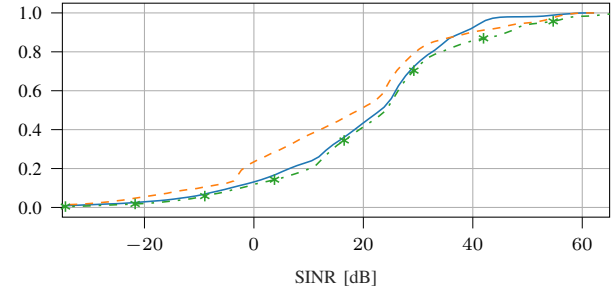}
\end{subfigure}
\caption{Comparison of the CDFs of different phased antenna array configurations at the UT side.}
\label{fig:array-size-comparison}
\end{figure*}

\section{Simulation Setup} 
\label{sec:simulation_setup}

We carried out a simulation campaign to evaluate the performance of different antenna and \gls{bf} configurations.
To this aim, we used the ns-3 \gls{mmwave} module extended with the proposed modeling framework.
The scenario we considered, depicted in \cref{fig:scenario}, models a parking lot with multiple cars (between 1.2 and 2.25~m high) and buildings, sampled every 5~ms.
Two \gls{mmwave} \glspl{bs} providing cellular coverage are placed on the front face of two buildings at a height of 3~m and are oriented with a bearing angle in the direction normal to the wall and with a downtilt of 12$^\circ$ with respect to the horizon.
Two users, UT\textsubscript{1} and UT\textsubscript{2}, both at a height of 1.5~m, are connected to the respective \gls{bs}.
During the simulation, UT\textsubscript{1} leaves the main building walking at 1.2~m/s up to point A and then starts driving towards the exit of the parking lot at 4.2~m/s, while UT\textsubscript{2} stands still at the center of the scenario.

The same type of \gls{bf} schemes is used by all nodes of the scenario.
We assume perfect channel knowledge for \gls{svd} \gls{bf}, computed for every received and transmitted packet.
Instead, to assess the impact of realistic mobility on this type of scenario, codebook-based \gls{bf} is only updated to find the best codeword pair for each TX/RX node pair every [10, 100, 1000]~ms.
Codebooks have been generated ensuring that adjacent beams cross at 3~dB below the maximum directivity and with no tapering across antennas.

The system operates at 28~GHz with a bandwidth of 400~MHz, and is configured with \gls{nr} numerology index~2.
The downlink traffic is generated by a remote server which transmits UDP packets to the users at a constant rate.
\cref{table:params} summarizes the parameters used in our evaluation.

To evaluate the communication performance, we considered both link-level and end-to-end metrics, including \gls{sinr} and \gls{snr} experienced by UT\textsubscript{1}, respectively showing the performance with and without the interference from the second cell, and APP-layer throughput.


\section{Simulation Results} 
\label{sec:simulation_results}

In this section, we present and comment the results obtained.
Unless explicitly stated, we consider the baseline simulation to have $4\times 4$ arrays for the \glspl{ut}, 3GPP antenna elements for the \glspl{bs}, and codebook-based \gls{bf} with 100~ms beam alignment, in addition to the parameters shown in \cref{table:params}.

In \cref{fig:temporal_evolution}, we reported the temporal evolution of the \gls{snr} and \gls{sinr} experienced by UT\textsubscript{1}. 
During the first part of the simulation, the \gls{snr} stays always above 50~dB and decreases as the user walks away, but the presence of interference strongly affects the channel quality, as shown by the behavior of the \gls{sinr}. 
At time instant A, the user gets in its car and starts driving towards the exit of the parking lot.
Shortly after 20~s and 30~s, some of the parked cars temporarily block the line of sight, making the channel quality suddenly drop.  
From time instant B to time instant C, both the \gls{snr} and the \gls{sinr} show an oscillating behavior caused by the presence of multiple reflections with similar path losses from the surrounding cars. 
The last part of the simulation is characterized by multiple blockage events due to the cars parked in the bottom part of the parking lot. 
During this phase, the \gls{snr} and \gls{sinr} exhibit similar behavior since the user is no longer subject to the inference caused by the communication between BS\textsubscript{2} and UT\textsubscript{2}.

\cref{fig:bf-comparison} shows the \glspl{cdf} of \gls{snr} and \gls{sinr} experienced by UT\textsubscript{1} with different \gls{bf} configurations.
We can notice that the \gls{svd} approach guarantees the best performance in terms of \gls{snr}, as supported by the theory, but not always when considering the \gls{sinr}, i.e., when interference is considered.
Since \gls{svd} \gls{bf} does not account for interference when computing the \gls{bf} vectors, while codebook \gls{bf} does so when probing the different codeword pairs, the performance gap between the two approaches is reduced and \gls{svd} may even be suboptimal, as shown in \cref{fig:bf-comparison-sinr}.
Moreover, it can be seen that the value of the refresh rate used to update the weight vectors affects the behavior of the codebook-based algorithm, providing better performance for more frequent updates.
Due to the geometry of the environment and the mobility, diminishing returns are clearly visible when reducing the beam alignment period from 100~ms to 10~ms making the extra overhead unnecessary.

\cref{fig:array-size-comparison} shows a comparison between different array sizes for the \glspl{ut}.
Clearly, the most complex configuration represented by a $4\times 4$ array is able to achieve the highest performance for both \gls{snr} and \gls{sinr}.
This is due both to the higher antenna gain obtained with the larger antenna array, but also to the reduced interference due to the higher directivity.
On the other hand, considering vertical 4 elements \glspl{ula} results in a very similar performance in the interference-free scenario, but vastly different performance when considering the interfering cell.
In fact, a vertical array is only able to produce directivity with cylindrical symmetry around the vertical axis.
Being both \glspl{bs} at the same height, a good \gls{bf} codeword able to improve the received power will also be likely to increase the downlink interference from the second cell.
On the other hand, when orienting the linear array horizontally, the cylindrical symmetry will also rotate over the horizontal axis.
In this case, the geometry of the environment and the positioning of the \glspl{bs} makes it less likely to incur strong interference.

\cref{fig:antenna-comparison} evaluates the impact of the element radiation pattern on the \gls{snr} experienced by the user.
Isotropic elements radiate equal power in all directions, and therefore provide a low directive gain but are able to cover a wide area.
On the contrary, elements characterized by the \gls{3gpp} pattern have high directivity but small beamwidth, which implies that the transmitted power is focused in a small portion of the space.
The best performance is achieved with the cosine pattern set to have a 3~dB beamwidth of 120$^\circ$, thus obtaining a maximum gain $G_{\rm max}=$5.7~dBi, as this represents a good compromise between directivity and beamwidth.

\cref{fig:fs} shows the average throughput achieved by UT\textsubscript{1} and UT\textsubscript{2} at the APP layer. 
With an inter-packet interval of 10~$\mu$s, the network is highly loaded and the scarcity of radio resources may prevent the recovery of the lost packets, e.g., by means of \gls{mac} and \gls{rlc} layer retransmissions. 
In this situation, the choice of the \gls{bf} algorithm may have a strong impact on the end-to-end performance, especially in the presence of user mobility. 
Indeed, as shown in \cref{fig:fs}, the higher channel gain provided by the \gls{svd}-based algorithm allows UT\textsubscript{1} to achieve higher throughput, while there is no benefit for UT\textsubscript{2} since it stays in the same position during the entire simulation.
Instead, with a higher inter-packet interval, the codebook-based algorithm achieves the same performance as the \gls{svd}, since the recovery mechanisms at the \gls{mac} and \gls{rlc} layers are able to compensate for the lower channel quality. 

\begin{figure}
  \centering
  \setlength\fwidth{\columnwidth}
  \setlength\fheight{.5\columnwidth}
  \input{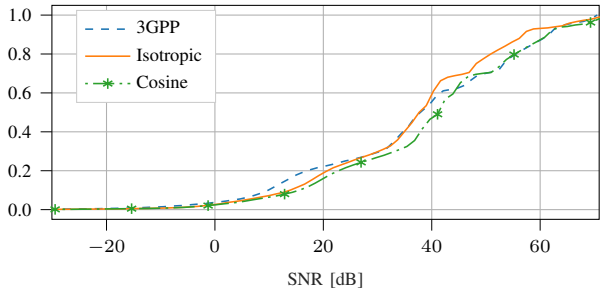}
\caption{Comparison of the CDFs of different antenna element patterns.}
\label{fig:antenna-comparison}
\end{figure}
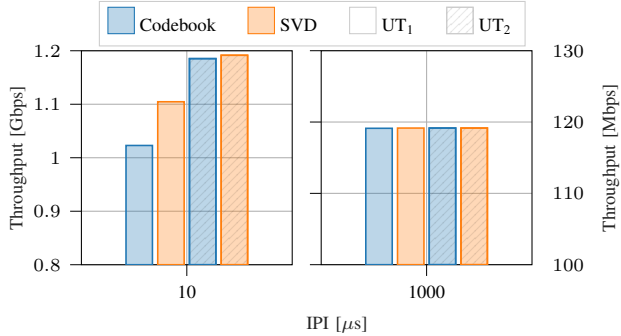
\begin{figure}
  \centering
  \setlength\fwidth{0.9\columnwidth}
  \setlength\fheight{.5\columnwidth}
\begin{tikzpicture}

\definecolor{color0}{rgb}{0.12156862745098,0.466666666666667,0.705882352941177}
\definecolor{color1}{rgb}{1,0.498039215686275,0.0549019607843137}
\definecolor{color2}{rgb}{0.172549019607843,0.627450980392157,0.172549019607843}
\definecolor{color3}{rgb}{0.83921568627451,0.152941176470588,0.156862745098039}
\definecolor{color4}{rgb}{0.580392156862745,0.403921568627451,0.741176470588235}
\definecolor{color5}{rgb}{0.549019607843137,0.337254901960784,0.294117647058824}
\definecolor{color6}{rgb}{0.890196078431372,0.466666666666667,0.76078431372549}

\pgfplotsset{every tick label/.append style={font=\scriptsize}}
\tikzstyle{every node}=[font=\scriptsize]

\begin{axis}[
  width=0.55\fwidth,
  height=\fheight,
  at={(0\fwidth,0\fheight)},
  every y tick scale label/.style={at={(rel axis cs:0.15,1.15)},anchor=north east},
tick align=outside,
tick pos=left,
x grid style={white!69.0196078431373!black},
xlabel={IPI [$\mu$s]},
ylabel={Throughput [Gbps]},
xmajorgrids,
xmin=0.5, xmax=1.5,
xtick style={color=black},
y grid style={white!69.0196078431373!black},
ymajorgrids,
ymin=0.8, ymax=1.2,
ytick style={color=black},
xtick={1},
xticklabels={10},
ybar,
xlabel style={xshift=2cm},
xlabel style={font=\scriptsize\color{white!15!black}},
ylabel style={font=\scriptsize\color{white!15!black}}
]
\addplot [semithick, color0, fill=color0, fill opacity=0.3]
table {%
1 1.0229386320984
};
\addplot [semithick, color1, fill=color1, fill opacity=0.3]
table {%
1 1.10467258923179
};

\addplot [semithick, color0, fill=color0, fill opacity=0.3, postaction={pattern=north east lines}]
table {%
1 1.1853003226044
};
\addplot [semithick, color1, fill=color1, fill opacity=0.3, postaction={pattern=north east lines}]
table {%
1 1.19157366487442
};

\end{axis}

\begin{axis}[
  width=0.55\fwidth,
  height=\fheight,
  at={(0.4\fwidth,0\fheight)},
tick align=outside,
tick pos=left,
x grid style={white!69.0196078431373!black},
ylabel={\scriptsize Throughput [Mbps]},
xmajorgrids,
xmin=1.5, xmax=2.5,
xtick style={color=black},
y grid style={white!69.0196078431373!black},
ymajorgrids,
ymin=100, ymax=130,
ytick style={color=black},
xtick={2},
xticklabels={1000},
ybar,
ylabel near ticks, yticklabel pos=right,
every y tick scale label/.style={at={(rel axis cs:0.88,1.15)},anchor=north west},
xlabel style={font=\scriptsize\color{white!15!black}},
ylabel style={font=\scriptsize\color{white!15!black}}
]
\addplot [semithick, color0, fill=color0, fill opacity=0.3]
table {%
2 119.122739927819
};
\addplot [semithick, color1, fill=color1, fill opacity=0.3]
table {%
2 119.156102231715
};

\addplot [semithick, color0, fill=color0, fill opacity=0.3, postaction={pattern=north east lines}]
table {%
2 119.157858142447
};
\addplot [semithick, color1, fill=color1, fill opacity=0.3, postaction={pattern=north east lines}]
table {%
2 119.157858142447
};

\end{axis}

\begin{axis}[
width=\fwidth,
height=\fheight,
at={(0\fwidth,0\fheight)},
legend cell align={left},
legend style={font=\scriptsize,at={(0.48,1.02)}, anchor=south, draw=white!80.0!black,
              /tikz/every even column/.append style={column sep=1em}},
tick pos=left,
x grid style={white!69.01960784313725!black},
xmajorgrids,
xmin=0, xmax=840,
xtick style={color=black},
y grid style={white!69.01960784313725!black},
ymajorgrids,
ymin=0, ymax=810,
ytick style={color=black},
hide y axis,
hide x axis,
legend columns=4,
xlabel style={font=\scriptsize\color{white!15!black}},
ylabel style={font=\scriptsize\color{white!15!black}}
]

\addlegendimage{ybar,ybar legend, color0, fill=color0, fill opacity=0.3, font=\footnotesize};
\addlegendentry{Codebook}

\addlegendimage{ybar,ybar legend, color1, fill=color1, fill opacity=0.3, font=\footnotesize};
\addlegendentry{SVD}

\addlegendimage{ybar,ybar legend,fill=white, fill opacity=0.3, font=\footnotesize};
\addlegendentry{UT\textsubscript{1}}

\addlegendimage{ybar,ybar legend,fill=white, fill opacity=0.3, postaction={pattern=north east lines},  font=\footnotesize};
\addlegendentry{UT\textsubscript{2}}
%
%
%
%
%
%

\end{axis}

\end{tikzpicture}
  \caption{APP-layer throughput for different inter-packet intervals for a UDP stream. \ml{svd? cb period?} \ml{forse meglio mettere due assi, sx in Gbps e dx in Mbps?}}
\label{fig:fs}
\end{figure}

\section{Conclusions} 
\label{sec:conclusions}

In this paper we presented a modeling framework for the end-to-end evaluation of 5G \gls{mmwave} cellular networks which is compliant with the \gls{3gpp}~\gls{nr} specifications. 
Our work extends the capabilities of the \gls{ns3} \gls{mmwave} module presented in~\cite{mezzavilla2018end} by providing (i) a ray-tracing based channel model for mobile users, which improves the spatio-temporal coherence over the previous stochastic channel~\cite{zugno2020implementation}, (ii) a flexible antenna module, comprising multiple parametric antenna elements as well as a generic interface for phased antenna arrays, and (iii) a \gls{bf} module supporting different algorithms for the computation of the optimal \gls{bf} vectors. 
Using this framework, we evaluated the performance achieved by different antenna configurations and \gls{bf} schemes in a realistic simulation scenario. 
Our results show that inaccurate antenna and \gls{bf} designs may provide sub-optimal channel gains and affect the performance of the higher layers. 
As future work, we plan to extend the proposed framework with a more realistic beam-management implementation, to account for the overhead introduced by beam search operations.

\bibliographystyle{IEEEtran}
\bibliography{bibl.bib}

\begin{thebibliography}{10}
\providecommand{\url}[1]{#1}
\csname url@samestyle\endcsname
\providecommand{\newblock}{\relax}
\providecommand{\bibinfo}[2]{#2}
\providecommand{\BIBentrySTDinterwordspacing}{\spaceskip=0pt\relax}
\providecommand{\BIBentryALTinterwordstretchfactor}{4}
\providecommand{\BIBentryALTinterwordspacing}{\spaceskip=\fontdimen2\font plus
\BIBentryALTinterwordstretchfactor\fontdimen3\font minus
  \fontdimen4\font\relax}
\providecommand{\BIBforeignlanguage}[2]{{%
\expandafter\ifx\csname l@#1\endcsname\relax
\typeout{** WARNING: IEEEtran.bst: No hyphenation pattern has been}%
\typeout{** loaded for the language `#1'. Using the pattern for}%
\typeout{** the default language instead.}%
\else
\language=\csname l@#1\endcsname
\fi
#2}}
\providecommand{\BIBdecl}{\relax}
\BIBdecl

\bibitem{38300}
3GPP, ``{NR and NG-RAN Overall Description},'' TS 38.300 (Rel. 15), 2018.

\bibitem{qualcomm201971}
------, ``New {WID} on extending current {NR} operation to 71 {GHz},''
  Qualcomm, RP-193229 - 3GPP TSG RAN Meeting 86, Dec. 2019.

\bibitem{rangan2014millimeter}
S.~Rangan, T.~S. Rappaport, and E.~Erkip, ``{Millimeter-wave Cellular Wireless
  Networks: Potentials and Challenges},'' \emph{Proceedings of the IEEE}, vol.
  102, no.~3, pp. 366--385, Mar. 2014.

\bibitem{va2015mmwaveCoherence}
V.~{Va} and R.~W. {Heath}, ``{Basic Relationship between Channel Coherence Time
  and Beamwidth in Vehicular Channels},'' in \emph{IEEE 82nd Vehicular
  Technology Conference (VTC2015-Fall)}, Boston, MA, USA, Sep. 2015.

\bibitem{giordani19beamManagement}
M.~{Giordani}, M.~{Polese}, A.~{Roy}, D.~{Castor}, and M.~{Zorzi}, ``{A
  Tutorial on Beam Management for 3GPP NR at mmWave Frequencies},'' \emph{IEEE
  Communications Surveys Tutorials}, vol.~21, no.~1, pp. 173--196, Sep. 2019.

\bibitem{zugno19simulation}
T.~Zugno, M.~Polese, M.~Lecci, and M.~Zorzi, ``{Simulation of Next-Generation
  Cellular Networks with Ns-3: Open Challenges and New Directions},'' in
  \emph{Workshop on Next-Generation Wireless with Ns-3 (WNGW)}, Florence,
  Italy, Jun. 2019.

\bibitem{zugno2020implementation}
T.~Zugno, M.~Polese, N.~Patriciello, B.~Bojovi\'{c}, S.~Lagen, and M.~Zorzi,
  ``{Implementation of a Spatial Channel Model for Ns-3},'' in \emph{Workshop
  on ns-3 (WNS3)}, Gaithersburg, MD, USA, Jun. 2020.

\bibitem{mezzavilla2018end}
M.~{Mezzavilla}, M.~{Zhang}, M.~{Polese}, R.~{Ford}, S.~{Dutta}, S.~{Rangan},
  and M.~{Zorzi}, ``{End-to-End Simulation of 5G mmWave Networks},'' \emph{IEEE
  Communications Surveys \& Tutorials}, vol.~20, no.~3, pp. 2237--2263, Third
  Quarter 2018.

\bibitem{8445856}
M.~{Polese} and M.~{Zorzi}, ``{Impact of Channel Models on the End-to-End
  Performance of Mmwave Cellular Networks},'' in \emph{{IEEE 19th International
  Workshop on Signal Processing Advances in Wireless Communications (SPAWC)}},
  Kalamata, Greece, Jun. 2018.

\bibitem{lecci2020fullStack}
M.~Lecci, P.~Testolina, M.~Polese, M.~Giordani, and M.~Zorzi, ``{Accuracy vs.
  Complexity for mmWave Ray-Tracing: A Full Stack Perspective},''
  https://arxiv.org/abs/2007.07125.

\bibitem{lecci20qdFormalization}
M.~Lecci, M.~Polese, C.~Lai, J.~Wang, C.~Gentile, N.~Golmie, and M.~Zorzi,
  ``{Quasi-Deterministic Channel Model for mmWaves: Mathematical Formalization
  and Validation},'' in \emph{IEEE Global Telecommunications Conference
  (GLOBECOM) (to be presented)}, Taipei, Taiwan, Dec. 2020.

\bibitem{qd-realization}
\BIBentryALTinterwordspacing
``{Quasi-deterministic mmWave Channel Simulator}.'' [Online]. Available:
  \url{https://github.com/wigig-tools/qd-realization}
\BIBentrySTDinterwordspacing

\bibitem{qd-channel}
\BIBentryALTinterwordspacing
``{ns-3 add-on for trace-based channel models}.'' [Online]. Available:
  \url{https://github.com/signetlabdei/qd-channel}
\BIBentrySTDinterwordspacing

\bibitem{balanis05antenna}
C.~A. Balanis, \emph{{Antenna Theory: Analysis and Design}}.\hskip 1em plus
  0.5em minus 0.4em\relax Wiley-Interscience, 2005.

\bibitem{rabbani2015improvement}
M.~S. Rabbani and H.~Ghafouri-Shiraz, ``{Size Improvement of Rectangular
  Microstrip Patch Antenna at Mmwave and Terahertz Frequencies},''
  \emph{Microwave and Optical Technology Letters}, vol.~57, no.~11, pp.
  2585--2589, Aug. 2015.

\bibitem{abdelkader2019systemlevel}
Y.~M. {Abdelkader}, M.~M. {Hamada}, and A.~N. {Mohieldin}, ``{System Level
  Co-Simulation Approach for Ultra-Wideband Massive MIMO Beam Forming Phased
  Array Transmitters},'' in \emph{31st International Conference on
  Microelectronics (ICM)}, Cairo, Egypt, Egypt, Mar. 2019.

\bibitem{8419177}
Y.~J. {Cho}, G.~{Suk}, B.~{Kim}, D.~K. {Kim}, and C.~{Chae}, ``{RF
  Lens-Embedded Antenna Array for mmWave MIMO: Design and Performance},''
  \emph{IEEE Communications Magazine}, vol.~56, no.~7, pp. 42--48, Jul. 2018.

\bibitem{menudier2018design}
C.~{Menudier}, J.~{Lintignat}, S.~{Mons}, P.~{M\'{e}drel}, N.~{Delhote},
  E.~{Ngoya}, S.~{Bila}, M.~{Thévenot}, B.~{Jarry}, P.~{Gamand}, J.~{Sombrin},
  and D.~{Baillargeat}, ``{Design and optimization of multielement antennas and
  RF circuits for beamforming with a reduced number of RF Front-ends},'' in
  \emph{IEEE MTT-S International Microwave Workshop Series on 5G Hardware and
  System Technologies (IMWS-5G)}, Dublin, Ireland, Aug. 2018.

\bibitem{3gpp.38.901}
3GPP, ``{Study on channel model for frequencies from 0.5 to 100 GHz},'' 3rd
  Generation Partnership Project (3GPP), Technical Report (TR) 38.901, Jun.
  2018, version 15.0.0.

\bibitem{zhang2017ns3}
M.~Zhang, M.~Polese, M.~Mezzavilla, S.~Rangan, and M.~Zorzi, ``{ns-3
  Implementation of the 3GPP MIMO Channel Model for Frequency Spectrum above 6
  GHz},'' in \emph{Workshop on ns-3 (WNS3)}, Porto, Portugal, Jun. 2017.

\end{thebibliography}

\end{document}